\documentclass[12pt]{article}

\usepackage{latexsym}
\usepackage{amssymb,amsfonts,amsmath}
\usepackage{graphicx} 
\usepackage{indentfirst}
\usepackage{bbm}
\usepackage{amssymb}
\usepackage{verbatim}
\usepackage{amsmath, amsthm,amssymb}
\usepackage{mathrsfs}
\usepackage{hyperref}
\usepackage{amsfonts}
\usepackage{dsfont}
\usepackage{cite}
\usepackage{xcolor}
\usepackage[multiple]{footmisc}

\parskip=\medskipamount

\parskip=\medskipamount

\usepackage[mathscr]{euscript}

\numberwithin{equation}{section}

\newcommand {\cN}{{\cal N}}

\newcommand {\cP}{{\cal P}}

\def\a{\alpha}
\def\b{\beta}

\def\d{\delta}

\def\g{\gamma}
\def\G{\Gamma}

\def\k{\kappa}

\def\m{\mu}
\def\n{\nu}
\def\o{\omega}

\def\q{\theta}
\def\r{\rho}
\def\s{\sigma}

\def\z{\zeta}
\def\D{\Delta}

\def\L{\Lambda}

\def\S{\Sigma}

\def\ri{{\rm i}}

\newcommand{\dmu}{{\dot{\mu}}}
\newcommand{\dnu}{{\dot{\nu}}}

\newcommand{\sSp}{\mathsf{Sp}}

\newcommand{\sSL}{\mathsf{SL}}
\newcommand{\sGL}{\mathsf{GL}}
\newcommand{\sSO}{\mathsf{SO}}

\newcommand{\sOSp}{\mathsf{OSp}}

\newcommand{\Iu}{\underline{I}}
\newcommand{\Au}{\underline{A}}

\newcommand{\alu}{{\underline{\a}}}

\newcommand{\Io}{{\overline{I}}}
\newcommand{\Ao}{\overline{A}}
\newcommand{\Bo}{\overline{B}}

\newcommand{\alo}{{\overline{\a}}}
\newcommand{\beo}{{\overline{\b}}}

\newcommand{\hal}{{\hat{\a}}}
\newcommand{\hbe}{{\hat{\b}}}
\newcommand{\hga}{{\hat{\g}}}
\newcommand{\hde}{{\hat{\d}}}

\newcommand{\ve}{\varepsilon}

\newcommand{\hf}{\frac12}

\newcommand{\be}{\begin{equation}}
\newcommand{\ee}{\end{equation}}
\newcommand{\bea}{\begin{eqnarray}}
\newcommand{\eea}{\end{eqnarray}}
\newcommand{\non}{\nonumber}
\newcommand{\ba}{\begin{array}}
\newcommand{\ea}{\end{array}}

\def\double #1{#1{\hbox{\kern-2pt $#1$}}}

\newcommand{\bsubeq}{\begin{subequations}}
\newcommand{\esubeq}{\end{subequations}}

\newcommand{\rd}{\mathrm d}
\begin{document}
\begin{titlepage}
\begin{flushright}
August, 2021\\
Revised version: October, 2021\\
\end{flushright}
\vspace{5mm}

\begin{center}
{\Large \bf 
Supertwistor realisations of AdS superspaces}
\end{center}

\begin{center}

{\bf
Sergei M. Kuzenko${}^a$ and Gabriele Tartaglino-Mazzucchelli${}^{b}$
} \\
\vspace{5mm}

\footnotesize{
${}^{a}${\it Department of Physics M013, The University of Western Australia,\\
35 Stirling Highway, Perth W.A. 6009, Australia}}  
~\\
\vspace{2mm}
\footnotesize{
${}^{b}${\it 
School of Mathematics and Physics, University of Queensland,
\\
 St Lucia, Brisbane, Queensland 4072, Australia}
}
\vspace{2mm}
~\\
\texttt{sergei.kuzenko@uwa.edu.au, 
g.tartaglino-mazzucchelli@uq.edu.au}\\
\vspace{2mm}

\end{center}

\begin{abstract}
\baselineskip=14pt
We propose supertwistor realisations of $(p,q)$ anti-de Sitter (AdS) superspaces in three dimensions and $\cal N$-extended AdS superspaces in four dimensions. For each superspace, we identify a two-point function that is invariant under the corresponding isometry supergroup. This two-point function is a supersymmetric extension (of a function) of the geodesic distance.
We also describe a bi-supertwistor formulation for $\cal N$-extended AdS superspace in four dimensions
and harmonic/projective extensions of $(p,q)$ AdS superspaces in three dimensions. 
\end{abstract}
\vspace{5mm}

\vfill
\end{titlepage}

\newpage
\renewcommand{\thefootnote}{\arabic{footnote}}
\setcounter{footnote}{0}

\tableofcontents{}
\vspace{1cm}
\bigskip\hrule


\allowdisplaybreaks

\section{Introduction} \label{section1}

Propagators in maximally symmetric spacetimes (see,  e.g., \cite{Avis:1977yn,Burgess:1984ti,Burges:1985qq,Allen:1985wd,Allen:1986qj,DHoker:1999bve} and references therein) make use of a unique two-point function which is invariant under the corresponding isometry group. Such a two-point function is readily constructed if one makes use of the well-known embedding formalisms for de Sitter and anti-de Sitter spaces. Off-shell supersymmetric field theories in AdS$_d$ are naturally formulated in appropriate AdS superspaces for $d\leq 5$. In order to develop quantum supergraph techniques in such a  superspace, it is useful to work with an embedding formalism. 
In this letter we propose  supertwistor formulations for the following superspace types:
(i)   $(p,q)$ anti-de Sitter (AdS) superspace in three dimensions;  and 
(ii) $\cN$-extended AdS superspace in four dimensions. 

Since the work by Ferber \cite{Ferber}, supertwistors have found numerous applications in  theoretical and mathematical physics. In particular, 
supertwistor realisations of compactified  $\cN$-extended Minkowski superspaces 
have been developed  in four
\cite{Manin,KNiederle} and three  \cite{Howe:1994ms,KPT-MvU}  dimensions and their harmonic/projective extensions  
\cite{Rosly2,LN,Howe:1994ms,HH1,HH2,K-compactified06, K-compactified12,KPT-MvU,BKS}.\footnote{Similar ideas  were applied in Ref. \cite{Kuzenko:2014yia} to develop
supertwistor realisations of the  $2n$-extended supersphere $S^{3|4n}$,
with $n=1,2,\dots$, as a homogeneous space of the three-dimensional
Euclidean superconformal group $\sOSp(2n|2,2)$.}
Recently, supertwistor formulations for conformal supergravity theories in diverse dimensions have been proposed \cite{HL1,HL2}.
Unlike in  Minkowski space,  not much is known about supertwistor realisations of AdS superspaces in diverse dimensions, to the best of our knowledge,
although (super)twistor descriptions of (super)particles in AdS have been studied in the literature 
\cite{CGKRZ,CRZ,CKR,BLPS,Zunger,Cederwall1,Cederwall2,AB-GT1,AB-GT2,Uvarov}.
Our goal in this paper is to fill the gap. Of course, for  theories in AdS it is always possible to use the standard coset space formalism, see, e.g., the famous Metsaev-Tseytlin construction of 
the type IIB superstring action in $\rm AdS_5 \times S^5$ \cite{Metsaev:1998it}.
However, manifest symmetry is one of the main virtues of (super)twistor techniques.

This paper is organised as follows. In section 2 we present the supertwistor realisations of $(p,q)$ AdS superspace in three 
dimensions. Section 3 is devoted to the four-dimensional $\cN$-extended case
which is then extended to a bi-supertwistor construction in section 4.
Section 5 is devoted to supertwistor constructions of harmonic/projective  AdS superspaces in three dimensions,
while section 6 contains concluding comments for our paper.
In the appendix we describe a supertwistor realisation 
of two-dimensional compactified 
Minkowski superspace
$\overline{\mathbb M}^{(2|p,q)}$.


\section{$(p,q)$ AdS superspace in three dimensions}\label{Section2}

The $(p,q)$ AdS superspaces in three dimensions (3D) were introduced in \cite{KLT-M12}
as backgrounds of the off-shell 3D $\cN$-extended conformal supergravity 
\cite{HIPT,KLT-M11} with  covariantly constant and Lorentz invariant torsion. 
In this paper we will restrict our attention to the conformally 
flat $(p,q)$ AdS superspaces\footnote{In the case $(p,q) = (\cN,0)$ there also exist non-conformally flat  
AdS superspaces if $\cN\geq 4$ \cite{KLT-M12}. They will be discussed elsewhere.} 
\bea
{\rm AdS}^{(3|p,q)} = \frac{ {\sOSp} (p|2; {\mathbb R} ) \times  {\sOSp} (q|2; {\mathbb R} ) } 
{ {\sSL}( 2, {\mathbb R}) \times {\sSO}(p) \times {\sSO}(q)}~,
\label{2.1}
\eea
which may be viewed as maximally supersymmetric solutions of $(p,q)$ AdS supergravity 
theories \cite{AT} (even though these theories are intrinsically formulated in components without auxiliary fields and can be recast in superspace only on the mass shell).\footnote{The coset spaces \eqref{2.1} were briefly discussed in \cite{BILS}.}
The superspaces \eqref{2.1} with $p+q \leq 4$ naturally originate as maximally  
supersymmetric solution of various off-shell AdS supergravity theories. In particular, 
AdS$^{(3|1,0)} $ corresponds to $\cN=1$ AdS supergravity \cite{GGRS}. 
The superspaces AdS$^{(3|1,1)} $ and AdS$^{(3|2,0)} $ correspond
to the off-shell formulations for $\cN=2$ AdS supergravity given in \cite{KLT-M11,KT-M11}.

As demonstrated in \cite{KLT-M12}, the isometry group of ${\rm AdS}^{(3|p,q)} $ is 
\bea
G= {\sOSp} (p|2; {\mathbb R} ) \times  {\sOSp} (q|2; {\mathbb R} )
\equiv G_{\rm L} \times G_{\rm R}~.
\label{2.2}
\eea
The same supergroup is also the superconformal group of compactified Minkowski superspace in two dimensions, 
$\overline{\mathbb M}^{(2|p,q)}$, with  its bosonic body being $\overline{\mathbb M}^2 = S^1 \times S^1$, the compactified two-dimensional  Minkowski space.\footnote{The supertwistor realisation of $\overline{\mathbb M}^{(2|p,q)}$ 
 is given in appendix \ref{AppendixA}.}
Our embedding formalism for ${\rm AdS}^{(3|p,q)} $ is constructed in terms of 2D supertwistors. 


\subsection{Algebraic background} 

We introduce two types of {\it pure} supertwistors, (i) a left supertwistor 
\bea
T_{\rm L} = (T_{\Ao}) =\left(
\begin{array}{c}
T_{\alo} \\
 T_{\Io}
\end{array}
\right)~, 
 \qquad \alo  = 1,2 ~, \quad \Io = 1, \dots, p ~;
 \label{2.3}
\eea
and (ii) a right supertwistor 
\bea
T_{\rm R} = (T_{\Au}) =\left(
\begin{array}{c}
T_\alu \\
 T_{\Iu}
\end{array}
\right)~, 
 \qquad \alu  = 1,2 ~, \quad \Iu = 1, \dots, q ~.
 \label{2.4}
\eea
In the case of even left supertwistors, $ T_\alo$ is bosonic
and $T_\Io$ is fermionic.
In the case of odd left supertwistors, $ T_\alo$ is fermionic while  $T^\Io$ is bosonic.
The even and odd left supertwistors are called pure.
We introduce the parity function $\ve ( T )$ defined as:
$\ve ( T ) = 0$ if $ T$ is even, and $\ve ( T ) =1$ if $T $ is odd.
Then the components $T_{\Ao}$ of a pure left supertwistor
 have the following  Grassmann parities
\bea
\ve ( T_{\Ao}) = \ve ( T ) + \ve_{\Ao} \quad (\mbox{mod 2})~,
\eea
where we have defined
\bea
 \ve_{\Ao} = \left\{
\begin{array}{c}
 0 \qquad \Ao=\alo \\
 1 \qquad \Ao=\Io
\end{array}
\right.{}~.
\non
\eea
Analogous definitions are introduced for the right supertwistors.

A pure left supertwistor is said to be real if its components obey the reality condition
\bea
(T_{\Ao})^* = (-1)^{\ve(T) \ve_{\Ao} + \ve_{\Ao}} \,T_{\Ao}~.
\label{realcon}
\eea
Real right supertwistors are similarly defined. 
The space of complex (real) even left supertwistors is naturally identified with
${\mathbb C}^{2|p}$ (${\mathbb R}^{2|p}$),
while the space of complex (real) odd left supertwistors may be identified with
${\mathbb C}^{p |2}$
(${\mathbb R}^{p |2}$).

We introduce  graded antisymmetric supermatrices ${\mathbb J}_{\rm L}$
and ${\mathbb J}_{\rm R}$ defined by 
\bea
{\mathbb J}_{\rm L} = ({\mathbb J}^{\Ao\, \Bo}) = \left(
\begin{array}{c ||c}
\ve_{\rm L}  ~&~ 0 \\\hline \hline
0 ~& ~{\rm i} \,{\mathbbm 1}_p
\end{array} \right) ~, \qquad
\ve_{\rm L}
=\big(\ve^{\alo  \beo} \big)
=\left(
\begin{array}{cc}
0  & { 1}\\
 -{ 1}  &    0
\end{array}
\right) ~,
\label{supermetric}
\eea
and similarly for ${\mathbb J}_{\rm R}$.
Here ${\mathbbm 1}_p $ denotes the unit  $p \times p $ matrix.
Associated with ${\mathbb J}_{\rm L} $ and ${\mathbb J}_{\rm R} $
are graded symplectic inner products on the spaces of pure left and right supertwistors, 
respectively.  For arbitrary pure left supertwistors $T$ and $S$,
their inner product is
\bea
\langle { T}| { S} \rangle_{{\mathbb J}_{\rm L} }: = {T}^{\rm sT}{\mathbb J}_{\rm L} {S}
~,
\label{innerp}
\eea
where the row vector  ${ T}^{\rm sT} $ is defined by
\bea
{ T}^{\rm sT} := \big( T_{\alo} , - (-1)^{\ve(T)}  T_{\Io} \big)
= (  T_{\Ao} (-1)^{\ve(T)\ve_{\Ao} +\ve_{\Ao}} )
\eea
and is called the super-transpose of $T$.
The above inner product is characterised by the symmetry property
\bea
\langle { T}_1 | { T}_2  \rangle_{{\mathbb J}_{\rm L}}
= -(-1)^{\ve(T_1)  \ve (T_2)} \langle { T}_2 | { T}_1  \rangle_{{\mathbb J}_{\rm L}} ~.
\eea
If $T_1$ and $T_2$ are real supertwistors, their inner product
obeys the reality relation
\bea
\Big( \langle { T}_1 | { T}_2  \rangle_{{\mathbb J}_{\rm L}} \Big)^* 
= - \langle { T}_2 | { T}_1  \rangle_{{\mathbb J}_{\rm L}}~.
\eea

We recall that the supergroup   $\sOSp(p|2; {\mathbb C})$
consists of those even $(2|p) \times (2|p)$ supermatrices
\bea
g = (g_{\Ao}{}^{\Bo}) ~, \qquad \ve(g_{\Ao}{}^{\Bo}) = \ve_{\Ao} + \ve_{\Bo} ~,
\eea
which preserve the inner product \eqref{innerp} under the action
\bea
T_{\rm L} =(T_{\Ao}) ~\to ~ g\cdot T_{\rm L} = (g_{\Ao}{}^{\Bo} T_{\Bo})~. 
\label{2.11}
\eea
Such a transformation maps the space of even (odd) supertwistors onto itself.
The condition of invariance of the inner  product \eqref{innerp}
under \eqref{2.11} is
\begin{subequations}\label{groupcond}
\bea
g^{\rm sT} {\mathbb J}_{\rm L} g = {\mathbb J}_{\rm L} ~, \qquad
\eea
where $g^{\rm sT}$ is the super-transpose of $g$ defined by
\bea
(g^{\rm sT})^{\Ao}{}_{\Bo} := (-1)^{\ve_{\Ao} \ve_{\Bo} + \ve_{\Bo}} g_{\Bo}{}^{\Ao}~.
\eea
\end{subequations}
The subgroup $G_{\rm L} \equiv \sOSp(p|2; {\mathbb R}) \subset \sOSp(p|2; {\mathbb C})$
consists of those transformations which preserve the reality condition
\eqref{realcon}, which means
\bea
\Big( g_{\Ao}{}^{\Bo} \Big)^*  = (-1)^{\ve_{\Ao} \ve_{\Bo} + \ve_{\Ao}} g_{\Ao}{}^{\Bo} 
\quad \Longleftrightarrow \quad g^\dagger = g^{\rm sT}~.
\eea
In conjunction with \eqref{groupcond}, this reality condition is equivalent to
\bea
g^\dagger {\mathbb J}_{\rm L} g = {\mathbb J}_{\rm L} ~.
\eea
Analogous definitions are introduced for the right supergroup
$G_{\rm R} \equiv \sOSp(q|2; {\mathbb R}) \subset \sOSp(q|2; {\mathbb C})$.


\subsection{Supertwistor realisation of $(p,q)$ AdS superspace}

In order to obtain a supertwistor realisation of $(p,q)$ AdS superspace, we introduce 
a space ${\mathfrak L}_{(p,q)}$. By definition, it  consists of all pairs 
$(\cP_{\rm L}, \cP_{\rm R})$, where 
\begin{subequations}
\bea
\cP_{\rm L} &=& (X_{\Ao}{}^\m ) ~, \qquad \m =1,2 
\eea
is a left real even two-plane, and 
\bea
\cP_{\rm R} &=& (Y_{\Au}{}^\m ) ~, \qquad \m =1,2 
\eea
\end{subequations}
is a  right real even two-plane, with the additional property 
\bea
\cP_{\rm L}^{\rm sT} {\mathbb J}_{\rm L} \cP_{\rm L} 
= \cP_{\rm R}^{\rm sT} {\mathbb J}_{\rm R} \cP_{\rm R}~.
\label{2.18}
\eea
A few comments are in order. The statement that $\cP_{\rm L}$ is even real, means that the two supertwistors $X_{\rm L}^\m$ are even and real. The property of $\cP_{\rm L}$ being a two-plane means that\footnote{More precisely, the body of the matrix $( X_{\alo}{}^\m )$ must be a nonsingular matrix. 
See \cite{Ideas} for the necessary information about
infinite dimensional Grassmann algebra $\L_\infty$ and supermatrices.}
\bea
\det ( X_{\alo}{}^\m ) \neq 0~.
\label{2.19}
\eea 
Similar statements hold for the right planes.
In the space ${\mathfrak L}_{(p,q)}$ we introduce the following equivalence relation 
\bea
(\cP_{\rm L}, \cP_{\rm R}) \sim (\cP_{\rm L} M, \cP_{\rm R}M)~, \qquad 
M \in \sGL( 2, {\mathbb R})~.
\label{2.20}
\eea

The supergroup \eqref{2.2}  acts on ${\mathfrak L}_{(p,q)}$ by the rule
\bea
(g_{\rm L}, g_{\rm R}) (\cP_{\rm L}, \cP_{\rm R}) := 
(g_{\rm L} \cP_{\rm L}, g_{\rm R} \cP_{\rm R}) ~,\qquad (g_{\rm L}, g_{\rm R}) \in 
{\sOSp} (p|2; {\mathbb R} ) \times  {\sOSp} (q|2; {\mathbb R} )~.
\label{2.21}
\eea
This action is naturally extended to the quotient space 
${\mathfrak L}_{(p,q)}/ \sim$. The latter proves to be a homogeneous space of 
${\sOSp} (p|2; {\mathbb R} ) \times  {\sOSp} (q|2; {\mathbb R} )$. It turns out that 
\bea
{\rm AdS}^{(3|p,q)} = {\mathfrak L}_{(p,q)}/ \sim~.
\label{2.22}
\eea

The equivalence relation \eqref{2.20} allows us to choose a gauge
\begin{subequations}
\bea
\cP_{\rm R}  =(Y_{\Au}{}^\m ) =\left(
\begin{array}{c}
\d_\alu{}^\m \\
 \ri \,\q_{\Iu}{}^\m 
\end{array}
\right)~, \qquad 
\cP_{\rm L}  =(X_{\Ao}{}^\m ) =\left(
\begin{array}{c}
x_\alo{}^\m \\
\ri \, \q_{\Io}{}^\m 
\end{array}
\right)~.
\label{2.23}
\eea
Then the condition \eqref{2.18} turns into 
\bea
x^{\rm T} \ve\, x = \ve - \ri \Big( \q_{\rm L}^{\rm T} \q_{\rm L} 
-\q_{\rm R}^{\rm T} \q_{\rm R} \Big)~.
\label{2.24}
\eea
\end{subequations}
This equation provides the embedding of ${\rm AdS}^{(3|p,q)} $ into ${\mathbb R}^{2,2|2p+2q}$. In the non-supersymmetric case, $p=q=0$, \eqref{2.24} is equivalent to 
\bea
x \in \sSp (2,{\mathbb R}) \cong \sSL (2,{\mathbb R}) ~,
\eea
which is the standard realisation of AdS$_3$.

Instead of using the gauge \eqref{2.23}, one can choose the alternative gauge condition 
\begin{subequations} \label{2.25}
\bea
\cP_{\rm L}  =(X_{\Ao}{}^\m ) =\left(
\begin{array}{c}
\d_\alo{}^\m \\
\ri \, \vartheta_{\Io}{}^\m 
\end{array}
\right)~,
 \qquad 
\cP_{\rm R}  =(Y_{\Au}{}^\m ) =\left(
\begin{array}{c}
y_\alu{}^\m \\
 \ri \,\vartheta_{\Iu}{}^\m 
\end{array}
\right)~.
\label{2.25a}
\eea
Then the condition \eqref{2.18} turns into 
\bea
y^{\rm T} \ve\, y = \ve - \ri \Big( \vartheta_{\rm R}^{\rm T} \vartheta_{\rm R} 
- \vartheta_{\rm L}^{\rm T} \vartheta_{\rm L} \Big)~.
\label{2.25b}
\eea
\end{subequations}


\subsection{$G$-invariant  two-point function on ${\rm AdS}^{(3|p,q)}$} 

Let $Z= (\cP_{\rm L}, \cP_{\rm R})$ and $\widetilde Z =(\widetilde{\cP}_{\rm L}, \widetilde{\cP}_{\rm R})$
be two points of ${\mathfrak L}_{(p,q)}$. We introduce the following two-point 
function\footnote{Due to the relations \eqref{2.18} and \eqref{2.19}, 
the combination $ \widetilde{\cP}_{\rm R}^{\rm sT} {\mathbb J}_{\rm R} \cP_{\rm R}$ 
is nonsingular.}
\bea
\o(Z,\widetilde Z) = \hf {\rm tr}\, \Big\{ \widetilde{\cP}_{\rm L}^{\rm sT} {\mathbb J}_{\rm L} \cP_{\rm L} 
\Big( \widetilde{\cP}_{\rm R}^{\rm sT} {\mathbb J}_{\rm R} \cP_{\rm R}\Big)^{-1}  \Big\}
~.
\label{2.26}
\eea
By construction, it is invariant under the group action \eqref{2.21}. 
The two-point function is also well defined on the quotient space \eqref{2.22}. 
Indeed, given two sets of equivalent points 
\bea
(\cP_{\rm L}, \cP_{\rm R}) \sim (\cP_{\rm L} M, \cP_{\rm R}M)~, \qquad
(\widetilde{\cP}_{\rm L}, \widetilde{\cP}_{\rm R}) 
\sim (\widetilde{\cP}_{\rm L} \widetilde{M}, \widetilde{\cP}_{\rm R} \widetilde{M})~, 
\eea
with $M , \widetilde{M} \in \sGL( 2, {\mathbb R})$, we have 
\bea
\widetilde{\cP}_{\rm L}^{\rm sT} {\mathbb J}_{\rm L} \cP_{\rm L} 
\sim \widetilde{M}^{\rm T} \widetilde{\cP}_{\rm L}^{\rm sT} {\mathbb J}_{\rm L} \cP_{\rm L} M~, \qquad 
\widetilde{\cP}_{\rm R}^{\rm sT} {\mathbb J}_{\rm R} \cP_{\rm R} 
\sim \widetilde{M}^{\rm T} \widetilde{\cP}_{\rm R}^{\rm sT} {\mathbb J}_{\rm R} \cP_{\rm R} M~,
\eea
and therefore the two-point function \eqref{2.26} does not change. 

It is instructive to evaluate  \eqref{2.26} in the non-supersymmetric case, $p=q=0$.
Assuming the gauge condition \eqref{2.23}, we then have 
\bea
x =\left(
\begin{array}{cc}
 x^0 +x^1  &  x^2 + x^3 \\
 x^2-x^3 &      x^0-x^1
\end{array}
\right) ~,\qquad (x^0)^2 +(x^3)^2 -(x^1)^2 -(x^2)^2 =1~,
\eea
and therefore 
\bea
w(x, \widetilde x) = \widetilde x^0 x^0 + \widetilde x^3 x^3
- \widetilde x^1 x^1- \widetilde x^2 x^2~.
\eea


\section{$\cN$-extended AdS superspace in four dimensions}\label{Section3}

The supergroup  $\sOSp(\cN|4; {\mathbb R})$ is 
the isometry group of four-dimensional $\cN$-extended 
AdS superspace
\bea
\rm{AdS}^{4|4\cN} =  \frac{\sOSp(\cN|4; {\mathbb R}) }{  \sSO(3,1) 
\times \sSO(\cN)} ~.
\label{3.1}
\eea
Here we describe a supertwistor realisation of this superspace. 
Our embedding formalism for ${\rm AdS}^{4|4\cN} $ is constructed in terms of 3D supertwistors. 

It should be pointed out that $\rm{AdS}^{4|4} $
 was introduced in \cite{Keck,Zumino77,IS}. It is a maximally supersymmetric solution of $\cN=1$ supergravity with a cosmological term, see \cite{GGRS,Ideas} for a review. The description of $\rm{AdS}^{4|8} $ as  a maximally supersymmetric solution of $\cN=2$ supergravity with a cosmological term was given in \cite{KLRT-M1,KT-M08,Butter:2011ym}. 
The conformal flatness of AdS$^{4|4}$ was established by Ivanov and Sorin  \cite{IS} and then reviewed in  textbooks \cite{GGRS,Ideas}. 
The superconformal flatness of $\rm{AdS}^{4|4\cN}$ was demonstrated in \cite{BILS}.
Ref. \cite{BKLT-M} described alternative  conformally flat realisations for 
AdS$^{4|4}$  and AdS$^{4|8}$ which are based on the use of Poincar\'e coordinates.


\subsection{Algebraic background} 

A supertwistor is a column vector
\bea
T = (T_A) =\left(
\begin{array}{c}
T_\hal \\
\hline \hline
 T_i
\end{array}
\right)~,
 \qquad
(T_\hal ) = \left(
\begin{array}{c}
 f_\a \\
  g^\b
  \end{array}
\right)~, \qquad \a, \b = 1,2 
\qquad i = 1, \dots, \cN ~.
\eea
Pure supertwistors are defined similarly to section \ref{Section2}.
Specifically the components $T_A$ of a pure supertwistor
 have the following  Grassmann parities
\bea
\ve ( T_A) = \ve ( T ) + \ve_A \quad (\mbox{mod 2})~,
\eea
where we have defined
\bea
 \ve_A = \left\{
\begin{array}{c}
 0 \qquad A=\hal \\
 1 \qquad A=i
\end{array}
\right.{}~.
\non
\eea

We choose the graded antisymmetric supermatrix
\bea
{\mathbb J} = ({\mathbb J}^{AB}) = \left(
\begin{array}{c ||c}
J ~&~ 0 \\\hline \hline
0 ~& ~{\rm i} \,{\mathbbm 1}_\cN
\end{array} \right) ~, \qquad
J
=\big(J^{\hat \a \hat \b} \big)
=\left(
\begin{array}{cc}
0  & {\mathbbm 1}_2\\
 -{\mathbbm 1}_2  &    0
\end{array}
\right) ~,
\label{supermetric2}
\eea
which allows us 
to define a graded symplectic inner product on the space of pure supertwistors by the rule: for arbitrary pure supertwistors $T$ and $S$,
the inner product is
\bea
\langle { T}| { S} \rangle_{\mathbb J}: = {T}^{\rm sT}{\mathbb J} \, {S}
~,
\label{innerp2}
\eea


\subsection{Supertwistor realisation of $\rm{AdS}^{4|4\cN} $}

We denote by ${\mathfrak E}_\cN$ the space of all real even supertwistors.
Next we introduce a complex frame in ${\mathfrak E}_\cN$  
\begin{subequations}\label{frame}
\bea
T^{\hat \m} = (T^\m , \bar T^{\dot \m} )~, \qquad 
T^\m = (T_A{}^\m ) ~,\qquad \bar T^{\dot \m} = (\bar T_A{}^{\dot \m }) ~.
\qquad \m, \dot \m =1,2~.
\label{3.6a} 
\eea
Here the supertwistor $\bar T^{\dot \m} $ is the complex conjugate of $T^\m$.
We require the elements of the frame to obey the conditions:
\bea
\ve_{\m\n} \langle  T^\m | T^\n \rangle_{\mathbb J} & \neq &0~; \\
 \langle  T^\m | \bar T^{\dot \n} \rangle_{\mathbb J} &= &0~.
 \label{3.6c}
 \eea
 \end{subequations}
We denote ${\mathfrak F}_\cN$ the space of all complex frames \eqref{frame}.

It is not difficult to construct explicit examples of complex frames \eqref{frame}.
Let $U^\m $ and $V^\m$ be real even supertwistors with the properties
\begin{subequations} 
\bea
\langle  U^\m | U^\n  \rangle_{\mathbb J} &=& \langle  V^\m | V^\n  \rangle_{\mathbb J} =0~,\\
\langle  U^\m | V^\n  \rangle_{\mathbb J} &=& - \langle  V^\m | U^\n  \rangle_{\mathbb J} 
= \d^{\m \n}~.
\eea
 \end{subequations}
 Such supertwistors originate as even vector-columns of an arbitrary  group element
 $g \in \sOSp(\cN|4; {\mathbb R})$. Then we define the complex even supertwistors
 \bea
 T^\m := U^\m + \ri \ve^{\m\s} V^\s~, 
 \qquad 
  \bar T^{\dot \m} := U^\m - \ri \ve^{\m\s} V^\s~,
  \eea
for which the properties \eqref{frame} hold. 
 
In the space of frames ${\mathfrak F}_\cN$, we introduce the following equivalence relation
\bea
T^\m \sim T^\n R_\n{}^\m ~, \qquad R \in \sGL(2,{\mathbb C})~.
\label{3.9}
\eea
The supergroup $ \sOSp(\cN|4; {\mathbb R})$ acts on ${\mathfrak F}_\cN$ by the rule 
\bea
g (T^\m , \bar T^{\dot \m} ) = (g T^\m , g  \bar T^{\dot \m} )~, 
\qquad g \in \sOSp(\cN|4; {\mathbb R})~.
\eea
This action is naturally extended to the quotient space 
${\mathfrak F}_\cN / \sim$. The latter proves to be a homogeneous space of 
$ \sOSp(\cN|4; {\mathbb R})$. It turns out that 
\bea
\rm{AdS}^{4|4\cN}= {\mathfrak F}_\cN/\sim~.
\label{3.11} 
\eea

\subsection{Anti-de Sitter space}

In order to prove \eqref{3.11}, it suffices to consider the non-supersymmetric case, 
$\cN=0$. Then we have 
\bea
T_{\hal}{}^{\hat \m}  T_{\hbe}{}^{\hat \n} 
T_{\hat \g}{}^{\hat \s}  T_{\hat \d}{}^{\hat \r} \ve_{\hat \m \hat \n \hat \s \hat \r} 
= \D \ve_{\hal \hbe \hga \hde}
= -\D\Big( J_{\hal \hbe} J_{\hga \hde} + J_{\hal \hga} J_{\hde \hbe} 
+ J_{\hal \hde} J_{\hbe \hga} \Big)~,
\label{3.12}
\eea
for some $\D\neq 0$. We know that 
\bea
 \langle  T^\m | T^\n \rangle_{\mathbb J} = \k \ve^{\m\n}~, \qquad 
 \langle  \bar T^{\dot \m} | \bar T^{\dot \n} \rangle_{\mathbb J} 
 =\bar \k \ve^{\dot \m \dot \n}~, 
 \label{3.13}
 \eea
 for some complex parameter $\k \neq 0$. Making use of \eqref{3.6a}, 
 \eqref{3.12} and \eqref{3.13}, we deduce that 
 \bea
 \bar \k T_{\hal}{}^\m T_{\hbe \m } 
 + \k \bar T_{\hal}{}^{\dot \m} \bar T_{\hbe \dot \m } = - \D J_{\hal \hbe} ~.
 \label{3.14}
 \eea
 It is useful to introduce the traceless part of the antisymmetric bi-twistor 
 $T_{\hal}{}^\m T_{\hbe \m }$,
 \bea
 T_{\langle \hal}{}^\m T_{\hbe \rangle \m } = T_{\hal}{}^\m T_{\hbe \m }
 - \hf J_{\hal \hbe} \k~, \qquad 
 J^{\hal \hbe}  T_{\langle \hal}{}^\m T_{\hbe \rangle \m } =0~.
 \eea
 Then the relation \eqref{3.14} is equivalent to the two identities:
 \begin{subequations} 
 \bea
 \bar \k T_{\langle \hal}{}^\m T_{\hbe \rangle  \m } 
 + \k \bar T_{\langle \hal}{}^{\dot \m} \bar T_{\hbe \rangle  \dot \m } &=&0~, 
 \label{3.16a}\\
 \D &=& -\k \bar \k~.
 \eea
 \end{subequations}
 Making use of the equivalence relation \eqref{3.9} allows us to choose a gauge 
\bea
\k = - \bar \k = \ri \ell ~, 
\label{3.17}
\eea
for a fixed real parameter $\ell$. Then \eqref{3.16a} turns into the reality condition 
\bea
T_{\langle \hal}{}^\m T_{\hbe \rangle  \m } 
= \bar T_{\langle \hal}{}^{\dot \m} \bar T_{\hbe \rangle  \dot \m } ~.
\eea

Associated with $T_{\langle \hal}{}^\m T_{\hbe \rangle  \m } $ is the real 5-vector 
\bea
X_{\hat a} := \hf (J\G_{\hat a} )^{\hal \hbe} T_{\langle \hal}{}^\m T_{\hbe \rangle  \m } 
=\hf (J\G_{\hat a} )^{\hal \hbe} T_{ \hal}{}^\m T_{\hbe   \m } 
~.
\eea
Here $\G_{\hat a}$ are real $4\times 4$ matrices which  obey the anti-commutation relations 
\bea
\{ \G_{\hat a}  , \G_{\hat b}  \} = 2\eta_{\hat a \hat b} {\mathbbm 1}_4~, \qquad 
\eta_{\hat a \hat b}= {\rm diag} \, (-+++-)~, \quad \hat a = 0,1,2,3,4 \equiv a, 3,4~.
\label{3.2}
\eea
These matrices constitute  a Majorana representation of the gamma-matrices
for pseudo-Euclidean space ${\mathbb R}^{3,2}$. The explicit realisation of 
$\G_{\hat a}$ is given, e.g., in \cite{KPT-MvU}.
Making use of the completeness relation
\bea
(J\, \G^{\hat a})^{\hat \a \hat \b} (J\, \G_{\hat a} )^{\hat \g \hat \d} 
= - J^{\hat \a \hat \b} J^{\hat \g \hat \d} 
+ 2 (J^{\hat \a \hat \g} J^{\hat \b \hat \d} - J^{\hat \a \hat \d} J^{\hat \b \hat \g} )~,
\eea 
we obtain 
\bea
X^{\hat a} X_{\hat a} = - \ell^2~.
\label{3.22}
\eea

The above twistor description of AdS$_4$ is equivalent to the bispinor formalism introduced in \cite{BFP}.


\subsection{$\sOSp(\cN|4; {\mathbb R})$-invariant  two-point function on 
${\rm AdS}^{4|4\cN}$} 

Let  $T^{\hat \m} $ and $\widetilde{T}^{\hat \m} $ be arbitrary points 
of ${\mathfrak F}_\cN$. The following two-point function 
\bea
\o( T, \widetilde T) := \frac{ \langle  \bar T^{\dot \m} | \widetilde{T}^\n \rangle_{\mathbb J}  
 \langle  \bar T_{\dot \m} | \widetilde{T}_\n \rangle_{\mathbb J}   } 
{ 
\langle  \bar T^{\dot \s} | \bar T_{\dot \s} \rangle_{\mathbb J} 
\langle  \widetilde{T}^\r | \widetilde{T}_\r \rangle_{\mathbb J} 
}
\label{3.23}
\eea
  is clearly $\sOSp(\cN|4; {\mathbb R})$-invariant. It is also invariant under equivalence transformations 
 \bea
 T^\m \to T^\n R_\n{}^\m~, \qquad   \widetilde{T}^\m \to \widetilde{T}^\n \widetilde{R}_\n{}^\m~, \qquad R, \widetilde{R} \in \sGL(2,{\mathbb C})~,
\eea
and therefore the two-point function is defined on the quotient space \eqref{3.11}.

In the non-supersymmetric case, $\cN=0$, \eqref{3.23}  is simply related to 
the AdS$_4$ two-point function $X^{\hat a} \widetilde{X}_{\hat a} $.
In the gauge \eqref{3.17}, we obtain 
\bea
X^{\hat a} \widetilde{X}_{\hat a} = - \ell^2 
+   \langle  \bar T^{\dot \m} | \widetilde{T}^\n \rangle_{\mathbb J}  
 \langle  \bar T_{\dot \m} | \widetilde{T}_\n \rangle_{\mathbb J} ~.
 \eea
 

\subsection{Poincar\'e coordinate patch in ${\rm AdS}^{4|4\cN}$} 

Let us consider an open subset of ${\rm AdS}^{4|4\cN}$ such that the upper
 $2\times 2$ 
block in 
\bea
T^\m =\left(
\begin{array}{c}
T_\a{}^\m 
\\
\hline
T^{\a \m} 
 \\
\hline \hline
 T_I{}^\m
\end{array}
\right)
\eea
is nonsingular. Then we can use the gauge freedom
\eqref{3.9} 
 to impose 
the condition \eqref{3.17} and choose $T_\a{}^\m \propto \d_\a{}^\m$.
Now, imposing the conditions \eqref{3.6c}, \eqref{3.13} and \eqref{3.17}, 
we obtain the general solution 
\bsubeq \label{3.27}
\bea
T^\m &=&\frac{1}{\sqrt{z_{(-)}} }\left(
\begin{array}{c}
\d_\a{}^\m
\\
\hline
\phantom{\Big|}
-x_{(-)}^{\b\m } +\frac{\ri}{2} ( \ell z_{(-)} +\q^2 ) \ve^{\b\m}
\phantom{\Big|}
\\
\hline \hline
\ri  \sqrt{2} \,\q_I{}^{\m}
\end{array}
\right)~,\\
\bar{T}^{\dot{\m}}
&=&
\frac{1}{\sqrt{z_{(+)}} }\left(
\begin{array}{c}
\d_\a{}^{\dot\m}
\\
\hline
\phantom{\Big|}
-x_{(+)}^{\b\dot\m } -\frac{\ri}{2} ( \ell z_{(+)} -\bar\q^2 ) \ve^{\b\dot{\m}}
\phantom{\Big|}
\\
\hline \hline
\ri  \sqrt{2} \,\bar\q_I{}^{\dot\m}
\end{array}
\right)~,
\eea
\esubeq
where we have denoted
\bsubeq
\bea
x_{(\pm)}^{\a\b } = x^{\a\b} \pm\ri \q_I{}^{(\a} \bar \q_I{}^{\b)} ~&,& \qquad
x^{\a\b}
=\left(
\begin{array}{cc}
x^0 -x^2   & -x^1 \\
-x^1  &    x^0 + x^2
\end{array}
\right) ~,
\\
z_{(\pm)}=z\pm\frac{1}{2\ell}(\q-\bar\q)^2~&,&\qquad
\q^2= \q_I{}^\a \q_{I\a} ~, \qquad \bar \q^2= \bar \q_I{}^\a \bar \q_{I\a} ~.
\eea
\esubeq
The real coordinates $z>0$ and $x^a = (x^0,x^1,x^2)$ parametrise AdS$_4$ in the Poincar\'e 
patch. They are related to the embedding coordinates $X^{\hat a} $, eq. \eqref{3.22}, 
as follows 
\bea
X^{\hat a} = (X^a, X^3, X^4) = \frac{1}{z} \Big( x^a\,, \frac{1-x^2 - (\ell z)^2 }{2}\, ,
 \frac{1+x^2 + (\ell z)^2 }{2}\Big)~, \quad x^2 =x^a x_a~.
\eea
In the non-supersymmetric case, $\cN=0$, the relations  \eqref{3.27} reduce to those given in 
\cite{BFP}.


\section{Bi-supertwistor construction for AdS$^{4|4\cN} $ }

Along with the supertwistor realisation of 
compactified  $\cN$-extended Minkowski superspaces in four dimensions, 
$\overline{\mathbb M}^{4|4\cN}$, there also exists the so-called bi-supertwistor
realisation for the same superspace which was introduced by Siegel
\cite{Siegel93,Siegel95} (see \cite{K-compactified12} for a modern description).
Here we describe its extension to  AdS$^{4|4\cN} $.

It should be mentioned that the bi-supertwistor construction 
of $\overline{\mathbb M}^{4|4\cN}$
was called ``superembedding formalism'' in
\cite{Goldberger:2011yp,Maio,Goldberger:2012xb}.
Indeed, this construction may be viewed as a specific example of
a general (super)embedding approach
reviewed in \cite{Sorokin} in application to superbranes.
This construction was advocated in \cite{Goldberger:2011yp,Maio,Goldberger:2012xb,Fitzpatrick:2014oza,Khandker:2014mpa}  
as a powerful alternative technique to compute correlation functions in conformal field theories, 
which is in a sense complementary to the more traditional 
superspace approaches pursued in \cite{Osborn,Park4,KT,Park3}.

Given a point in  ${\mathfrak F}_\cN$, we associate with it the  graded antisymmetric matrices
\begin{subequations}\label{4.1}
\bea
X_{AB} &:=& -2 \frac{ T_A{}^{ \m}  {T}_{B\m}  } 
{ \langle   T^{ \n} |  T_{\n} \rangle_{\mathbb J} }
=- (-1)^{\ve_A \ve_B}  X_{BA}~,
\\
\bar X_{AB} &:=& -2 \frac{ \bar T_A{}^{ \dmu}  \bar T_{B\dmu}    } 
{ \langle   \bar T^{ \dnu} |  \bar T_{\dnu} \rangle_{\mathbb J} }
=- (-1)^{\ve_A \ve_B}  \bar X_{BA}~.
\eea
\end{subequations} 
These supermatrices are invariant under arbitrary equivalence transformations
\bea
T^\m \to T^\n R_\n{}^\m ~, \qquad R \in \sGL(2,{\mathbb C})~,
\eea
and therefore they may be used to parametrise AdS$^{4|4\cN} $.
The bi-supertwistors \eqref{4.1} have the following properties: 
\begin{subequations} \label{4.3}
\bea
X_{[AB} X_{CD \}} &=&0~, \\
(-1)^{\ve_B} X_{AB} {\mathbb J}^{BC} X_{CD} &=& X_{AD} ~,\\
 {\mathbb J}^{BA}  X_{AB}&=& 2~, \\
(-1)^{\ve_B} X_{AB} {\mathbb J}^{BC} \bar X_{CD} &=& 0~.
\eea
\end{subequations}
Making use of the results of  \cite{K-compactified12}, the bi-supertwistor formulation 
for AdS$^{4|4\cN}$ defined by \eqref{4.3} may be shown to be equivalent to the supertwistor one described in section \ref{Section3}.


\section{Harmonic/projective AdS superspaces}

The supertwistor realisations of AdS$^{(3|p,q)}$ and AdS$^{4|4\cN}$, which have been described in sections \ref{Section2} and \ref{Section3}, make use of even supertwistors. In order to formulate AdS analogues of the harmonic \cite{GIKOS,GIOS} and projective \cite{KLR,LR-projective,LR-projective2} superspaces, odd supertwistors must be taken into account. The corresponding technical details are analogous to the 3D and 4D flat-superspace constructions of Refs. \cite{KPT-MvU,K-compactified06}
which built on earlier works \cite{Rosly2,LN,HH2}. This is why we provide such AdS formulations only in three dimensions.

Here we consider particular members of the family of 3D $(p,q)$ AdS superspaces,
specifically ${\rm AdS}^{(3|\cN,0)} \equiv {\rm AdS}^{3|2\cN}$. 
For a fixed $\cN =p+q$, the specific feature of  ${\rm AdS}^{(3|\cN,0)} $ 
and  ${\rm AdS}^{(3|0,\cN)} $
is that 
 the corresponding $R$-symmetry subgroup of the isometry group \eqref{2.2} is maximal and coincides with the $R$-symmetry subgroup of the $\cN$-extended superconformal group $\sOSp(\cN|4; {\mathbb R})$, which is  $\sSO(\cN)$.\footnote{The superspaces 
${\rm AdS}^{(3|\cN,0)} $ 
and  ${\rm AdS}^{(3|0,\cN)} $
  are related to each other by a parity transformation.}
Superspace $ {\rm AdS}^{3|2\cN}$ can be extended to $ {\rm AdS}^{3|2\cN} \times {\mathbb X}^\cN_1$,
where the internal space ${\mathbb X}^\cN_1$ is realised in terms of left complex {\it odd} supertwistors\footnote{One can also consider superspaces 
 $ {\rm AdS}^{3|2\cN} \times {\mathbb X}^\cN_m$, for any integer
$m\leq [\cN/2]$, with $[\cN/2]$ being 
the integer part of $\cN/2$. 
Space ${\mathbb X}^\cN_m$ is realised in terms of $m$ left odd  complex supertwistors
${\S}^{i}$, with $i=1,\dots, m$, such that  (i) the bodies of ${ \S}^{i}$ are linearly independent; 
(ii) the $\S^i$  obey the constraints 
$\cP_{\rm L}^{\rm sT} {\mathbb J}_{\rm L}  \S_{\rm L}^i =0$
and
$\S^i_{\rm L}{}^{\rm sT} {\mathbb J}_{\rm L} \S^j_{\rm L} =0$; 
and (iii) the $\S^i$  are defined modulo 
the equivalence relation $\S^i \sim \S^j D_j{}^i $, with $D \in   \sGL(m,{\mathbb C})$.}
\bea
\S_{\rm L} = \left( \begin{array}{c}
 \r_{\alo}  \\  \z_{\Io} 
\end{array} 
\right)~,\qquad  \z_{\Io} \neq 0~,
\eea
which are subject to the constraints 
\bea
\cP_{\rm L}^{\rm sT} {\mathbb J}_{\rm L}  \S_{\rm L} =0~, 
\qquad \
\S_{\rm L}^{\rm sT} {\mathbb J}_{\rm L} \S_{\rm L} =0~,
\eea
and are defined modulo the equivalence relation
\bea
\S_{\rm L } \sim c \, \S_{\rm L} ~,
 \qquad c\in {\mathbb C} \setminus  \{0\}~.
 \eea
 In the gauge \eqref{2.25}, the above constraints take the form:
  \bea
  \r_{\alo} = \z_{\Io} \vartheta_{\Io}{}^{\beo}\ve_{\beo \alo}~, \qquad 
  \z_{\Io} \z_{\Io} = \ri \r_{\alo} \ve^{\alo \beo} \r_{\beo}~.
  \eea
For $\cN>2$ the internal manifold  ${\mathbb X}^\cN_1$ proves to be a symmetric 
space, 
\bea
{\mathbb X}^\cN_1= \frac{\sSO (\cN) }{ \sSO (\cN-2) \times \sSO(2) }~, \qquad \cN>2~.
\eea
In the $\cN=3$ case, the internal space ${\mathbb X}^3_1$ is ${\mathbb C}P^1$, 
while for $\cN=4$ one obtains ${\mathbb X}^4_1 ={\mathbb C}P^1 \times {\mathbb C}P^1$, see  \cite{KPT-MvU} for the details. 

It is obvious that the above construction naturally extends to the case of $(p,q)$ AdS superspaces with $p\geq q>0$. Technical details will be skipped.


\section{Conclusion}

In this paper we have presented supersymmetric extensions of the twistor descriptions 
 of AdS$_3$ and AdS$_4$.
Specifically, we have proposed supertwistor realisations of $(p,q)$ AdS superspaces in three dimensions and $\cal N$-extended AdS superspaces in four dimensions. In the three-dimensional case, we have also presented
harmonic/projective superspace formulations of $(p,q)$ AdS supersymmetry, and 
these results can be readily extended to four dimensions.

One of the main results of our paper is the construction of manifestly supersymmetric two-point functions in AdS$^{(3|p,q)}$ and AdS$^{4|4\cN}$.
In Minkowski backgrounds, the embedding approach is known to be
a powerful framework for deciphering the structure of correlation functions in conformal field theories -- see, e.g.,
\cite{Costa:2011mg,Goldberger:2011yp,Maio,Goldberger:2012xb,Fitzpatrick:2014oza,Khandker:2014mpa}.
Analogously, it is of interest for several applications to study $n$-point correlation functions in AdS by employing symmetry arguments, see, e.g., \cite{BFP} and references therein for a recent discussion  in the non-supersymmetric case.
The results of our work open new avenues to perform manifestly supersymmetric
studies of correlation functions in AdS$_3$ and AdS$_4$. 
We aim to look into this direction in the near future.


\noindent
{\bf Acknowledgements:}\\
We are grateful to Michael Ponds for comments on the manuscript.
The work of SMK is supported in part by the Australian 
Research Council, project No. DP200101944.
The work of GT-M is supported by the Australian Research Council (ARC)
Future Fellowship FT180100353, and by the Capacity Building Package of the University
of Queensland.

\appendix

\section{Compactified $(p,q)$ Minkowski superspace in two dimensions}
\label{AppendixA}

For completeness, in this appendix we describe a supertwistor realisation 
of 2D compactified Minkowski superspace
$\overline{\mathbb M}^{(2|p,q)}$. This superspace will be identified with 
\bea
\overline{\mathbb M}^{(2|p,q)} = \L_{(p,q)} /\sim~.
\label{A.1}
\eea
Here $ \L_{(p,q)} $ is the space of  real even supertwistor pairs $(T_{\rm L} , T_{\rm R})$, 
where  $T_{\rm L} $  and $T_{\rm R}$ are left and right even real  supertwistors
of the form \eqref{2.3} and \eqref{2.4}, respectively, with non-zero bosonic parts, 
\bea 
{\mathfrak T}_{\rm L} := ( T_{\alo})  \neq 0~, \qquad 
{\mathfrak T}_{\rm R} := ( T_{\alu})  \neq 0~.
\eea
The equivalence relation in \eqref{A.1} is defined by 
\bea
(T_{\rm L} , T_{\rm R}) \sim ( \r_{\rm L} T_{\rm L} , \r_{\rm R} T_{\rm R})~, \qquad 
 \r_{\rm L} ,  \r_{\rm R} \in {\mathbb R} - \{ 0\}~.
 \label{A.3}
 \eea
The supergroup \eqref{2.2}  acts on $\L_{(p,q)}$ by the rule
\bea
(g_{\rm L}, g_{\rm R}) (T_{\rm L}, T_{\rm R}) := 
(g_{\rm L} T_{\rm L}, g_{\rm R} T_{\rm R}) ~,\qquad (g_{\rm L}, g_{\rm R}) \in 
{\sOSp} (p|2; {\mathbb R} ) \times  {\sOSp} (q|2; {\mathbb R} )~.
\eea
This action is naturally extended to the quotient space 
$\L_{(p,q)}/ \sim$. The latter proves to be a homogeneous space of 
${\sOSp} (p|2; {\mathbb R} ) \times  {\sOSp} (q|2; {\mathbb R} )$. 

Let us define one-forms
\bea
\o_{\rm L} = - T_{\rm L}^{\rm sT} {\mathbb J}_{\rm L} \rd T_{\rm L} ~, \qquad
\o_{\rm R} = - T_{\rm R}^{\rm sT} {\mathbb J}_{\rm R} \rd T_{\rm R}~.
\eea
They  have the following properties: (i) $\o_{\rm L} $ and $\o_{\rm R} $
 are invariant under the action of ${\sOSp} (p|2; {\mathbb R} ) \times  {\sOSp} (q|2; {\mathbb R} )$; and (ii)  $\o_{\rm L} $ and $\o_{\rm R} $ scale under point-dependent 
 (local)  equivalence transformations, 
 \bea
  \o_{\rm L} \to \r_{\rm L}^2 \o_{\rm L} ~, 
  \qquad   \o_{\rm R} \to \r_{\rm R}^2 \o_{\rm R} ~.
  \label{A.5}
  \eea
 Therefore we can define a superconformal metric on $\overline{\mathbb M}^{(2|p,q)} $
 by the rule 
 \bea
 \rd s^2 =  \o_{\rm L}  \o_{\rm R} ~.
 \label{metric}
 \eea

In order to get a better feeling for the above construction, let us consider 
the non-supersymmetric case, $p=q=0$. The elements of $\L=\L_{(0,0)}$ are all possible pairs $(T_{\rm L} , T_{\rm R}) = (T_{\alo} , T_{\alu})$, where the real two-component spinors $T_{\alo} $ and $T_{\alu}$ are non-zero. 
The freedom to perform equivalence transformations \eqref{A.3} 
can be partially fixed by imposing the conditions
\bea
(T_{\bar 1} )^2 + (T_{\bar 2})^2 =1~, \qquad 
(T_{\underline 1} )^2 + (T_{\underline 2})^2 =1~.
\label{A.7}
\eea
In this gauge, the equivalence relation \eqref{A.3} reduces to $T_{\alo} \sim -T_{\alo} $
and $T_{\alu} \sim -T_{\alu} $. It is seen that the quotient space $\L/\sim $ is $S^1 \times S^1$.

Instead of imposing the conditions \eqref{A.7}, we can introduce inhomogeneous 
(North-chart) coordinates for the one-spheres, 
\bea
T_{\rm L} = \left(
\begin{array}{c}
 x_{\rm L} \\
  1
  \end{array}
\right)~, \qquad 
 T_{\rm R} = \left(
\begin{array}{c}
 x_{\rm R} \\
 1
  \end{array}
\right)~.
\label{A.8}
\eea
Then the one-forms \eqref{A.5} take the form 
\bea
 \o_{\rm L} = \rd x_{\rm L} ~, 
  \qquad   \o_{\rm R} =\rd x_{\rm R} ~,
  \eea
  and the metric \eqref{metric} becomes $ \rd s^2 =  x_{\rm L}  x_{\rm R} $.
Given a group element 
 \bea
g_{\rm L}
=\left(\begin{array}{cc}
a  & b\\
c  &    d
\end{array}
\right) 
  \in G_{\rm L}= \sSp (2,{\mathbb R}) \cong \sSL (2,{\mathbb R}) ~,
\eea
it acts on $T_{\rm L}$, eq. \eqref{A.8}, by the fractional linear transformation
\bea
x_{\rm L} \to \frac{a x_{\rm L} +b}{cx_{\rm L} +d} \quad \implies \quad
\rd x_{\rm L} \to \frac{\rd x_{\rm L} }{(cx_{\rm L} +d)^2} ~.
\eea
Given a group element $g_{\rm R}  \in G_{\rm R}= \sSp (2,{\mathbb R}) $, it generates a similar fractional linear transformation of $x_{\rm R}$.  Under the action of 
$(g_{\rm L}, g_{\rm R}) \in G_{\rm L} \times G_{\rm R}$, 
the metric
$ \rd s^2 =  x_{\rm L}  x_{\rm R} $ scales.

\begin{footnotesize}

\end{footnotesize}


\begin{thebibliography}{66}

\bibitem{Avis:1977yn}
S.~J.~Avis, C.~J.~Isham and D.~Storey,
``Quantum field theory in anti-de Sitter space-time,''
Phys. Rev. D \textbf{18}, 3565 (1978).

\bibitem{Burgess:1984ti}
C.~P.~Burgess and C.~A.~L\"utken,
``Propagators and effective potentials in anti-de Sitter space,''
Phys. Lett. B \textbf{153}, 137-141 (1985).

\bibitem{Burges:1985qq}
C.~J.~C.~Burges, D.~Z.~Freedman, S.~Davis and G.~W.~Gibbons,
``Supersymmetry in anti-de Sitter space,''
Annals Phys. \textbf{167}, 285 (1986).

\bibitem{Allen:1985wd}
B.~Allen and T.~Jacobson,
``Vector two-point functions in maximally symmetric spaces,''
Commun. Math. Phys. \textbf{103}, 669 (1986).

\bibitem{Allen:1986qj}
B.~Allen and C.~A.~L\"utken,
``Spinor two-point functions in maximally symmetric spaces,''
Commun. Math. Phys. \textbf{106}, 201 (1986)
doi:10.1007/BF01454972

\bibitem{DHoker:1999bve}
E.~D'Hoker, D.~Z.~Freedman, S.~D.~Mathur, A.~Matusis and L.~Rastelli,
``Graviton and gauge boson propagators in AdS(d+1),''
Nucl. Phys. B \textbf{562}, 330 (1999)
[arXiv:hep-th/9902042 [hep-th]].

\bibitem{Ferber}
  A.~Ferber, ``Supertwistors and conformal supersymmetry,''
  Nucl.\ Phys.\ B {\bf 132}, 55 (1978).


\bibitem{Manin} Yu. I. Manin, 
``Holomorphic supergeometry and Yang-Mills superfields,''
J. Sov. Math. {\bf 30}, 1927 (1985);
{\it Gauge Field Theory and Complex Geometry},
Springer, Berlin, 1988.

\bibitem{KNiederle}
  M.~Kotrla and J.~Niederle,
  ``Supertwistors and superspace,''
  Czech.\ J.\ Phys.\ B {\bf 35}, 602 (1985).

\bibitem{Howe:1994ms} 
  P.~S.~Howe and M.~I.~Leeming,
  ``Harmonic superspaces in low dimensions,''
  Class.\ Quant.\ Grav.\  {\bf 11}, 2843 (1994)
  [hep-th/9408062].




\bibitem{KPT-MvU}
S.~M.~Kuzenko, J.-H.~Park, G.~Tartaglino-Mazzucchelli and R.~Unge,
``Off-shell superconformal nonlinear sigma-models in three dimensions,''
JHEP {\bf 1101}, 146 (2011)
  [arXiv:1011.5727 [hep-th]].


\bibitem{Rosly2}
  A.~A.~Rosly,
 ``Gauge fields in superspace and twistors,''
  Class.\ Quant.\ Grav.\  {\bf 2}, 693 (1985).

\bibitem{LN}
  J.~Lukierski and A.~Nowicki,
  ``General superspaces from supertwistors,''
  Phys.\ Lett.\ B {\bf 211}, 276 (1988).
  
    
 \bibitem{HH1}
G.~G.~Hartwell and P.~S.~Howe,
``(N, p, q) harmonic superspace,''
Int. J. Mod. Phys. A \textbf{10}, 3901-3920 (1995)
[arXiv:hep-th/9412147 [hep-th]]. 

\bibitem{HH2}
  P.~S.~Howe and G.~G.~Hartwell,
  ``A superspace survey,''
  Class.\ Quant.\ Grav.\  {\bf 12}, 1823 (1995).



\bibitem{K-compactified06}
 S.~M.~Kuzenko,
``On compactified harmonic/projective superspace, 5D superconformal
theories, and all that,'' Nucl.\ Phys.\  B {\bf 745}, 176 (2006)
[arXiv:hep-th/0601177].

\bibitem{K-compactified12}
S.~M.~Kuzenko,
``Conformally compactified Minkowski superspaces revisited,''
JHEP {\bf 1210}, 135 (2012) [arXiv:1206.3940 [hep-th]].


\bibitem{BKS}
E.~I.~Buchbinder, S.~M.~Kuzenko and I.~B.~Samsonov,
``Superconformal field theory in three dimensions: Correlation functions of conserved currents,''
JHEP \textbf{06} (2015), 138
[arXiv:1503.04961 [hep-th]].  

\bibitem{Kuzenko:2014yia}
S.~M.~Kuzenko and D.~Sorokin,
``Superconformal structures on the three-sphere,''
JHEP \textbf{10}, 080 (2014)
[arXiv:1406.7090 [hep-th]].


\bibitem{HL1}
P.~S.~Howe and U.~Lindstr\"om,
``Local supertwistors and conformal supergravity in six dimensions,''
Proc. Roy. Soc. Lond. A \textbf{476}, no.2243, 20200683 (2020)
[arXiv:2008.10302 [hep-th]].


\bibitem{HL2}
P.~S.~Howe and U.~Lindstr\"om,
``Superconformal geometries and local twistors,''
JHEP \textbf{04}, 140 (2021)
[arXiv:2012.03282 [hep-th]].

\bibitem{CGKRZ}
P.~Claus, M.~Gunaydin, R.~Kallosh, J.~Rahmfeld and Y.~Zunger,
``Supertwistors as quarks of $SU(2, 2 | 4)$,''
JHEP \textbf{05}, 019 (1999)
[arXiv:hep-th/9905112 [hep-th]].

\bibitem{CRZ}
P.~Claus, J.~Rahmfeld and Y.~Zunger,
``A simple particle action from a twistor parametrization of AdS(5),''
Phys. Lett. B \textbf{466}, 181-189 (1999)
[arXiv:hep-th/9906118 [hep-th]].

\bibitem{CKR}
P.~Claus, R.~Kallosh and J.~Rahmfeld,
``BRST quantization of a particle in AdS(5),''
Phys. Lett. B \textbf{462}, 285-293 (1999)
[arXiv:hep-th/9906195 [hep-th]].

\bibitem{BLPS}
I.~A.~Bandos, J.~Lukierski, C.~Preitschopf and D.~P.~Sorokin,
``OSp supergroup manifolds, superparticles and supertwistors,''
Phys. Rev. D \textbf{61}, 065009 (2000)
[arXiv:hep-th/9907113 [hep-th]].

\bibitem{Zunger}
Y.~Zunger,
``Twistors and actions on coset manifolds,''
Phys. Rev. D \textbf{62}, 024030 (2000)
[arXiv:hep-th/0001072 [hep-th]].

\bibitem{Cederwall1}
M.~Cederwall,
``Geometric construction of AdS twistors,''
Phys. Lett. B \textbf{483}, 257-263 (2000)
[arXiv:hep-th/0002216 [hep-th]].

\bibitem{Cederwall2}
M.~Cederwall,
``AdS twistors for higher spin theory,''
AIP Conf. Proc. \textbf{767}, no.1, 96-105 (2005)
[arXiv:hep-th/0412222 [hep-th]].

\bibitem{AB-GT1}
A.~S.~Arvanitakis, A.~E.~Barns-Graham and P.~K.~Townsend,
``Anti-de Sitter particles and manifest (super)isometries,''
Phys. Rev. Lett. \textbf{118}, no.14, 141601 (2017)
[arXiv:1608.04380 [hep-th]].

\bibitem{AB-GT2}
A.~S.~Arvanitakis, A.~E.~Barns-Graham and P.~K.~Townsend,
``Twistor description of spinning particles in AdS,''
JHEP \textbf{01}, 059 (2018)
[arXiv:1710.09557 [hep-th]].


\bibitem{Uvarov}
D.~V.~Uvarov,
``Supertwistor formulation for massless superparticle in $AdS_5\times S^5$ superspace,''
Nucl. Phys. B \textbf{936}, 690-713 (2018)
[arXiv:1807.08318 [hep-th]].

\bibitem{Metsaev:1998it}
R.~R.~Metsaev and A.~A.~Tseytlin,
``Type IIB superstring action in AdS(5) x S**5 background,''
Nucl. Phys. B \textbf{533}, 109-126 (1998)
[arXiv:hep-th/9805028 [hep-th]].

\bibitem{KLT-M12} 
S.~M.~Kuzenko, U.~Lindstr\"om and G.~Tartaglino-Mazzucchelli,
``Three-dimensional (p,q) AdS superspaces and matter couplings,''
JHEP {\bf 1208}, 024 (2012)  
[arXiv:1205.4622 [hep-th]].


\bibitem{HIPT}
  P.~S.~Howe, J.~M.~Izquierdo, G.~Papadopoulos and P.~K.~Townsend,
  ``New supergravities with central charges and Killing spinors in 2+1 dimensions,''
  Nucl.\ Phys.\  B {\bf 467}, 183 (1996)
  [arXiv:hep-th/9505032].


\bibitem{KLT-M11}
S.~M.~Kuzenko, U.~Lindstr\"om and G.~Tartaglino-Mazzucchelli,
  ``Off-shell supergravity-matter couplings in three dimensions,''
  JHEP {\bf 1103}, 120 (2011)
  [arXiv:1101.4013 [hep-th]].


\bibitem{AT}
  A.~Ach\'ucarro and P.~K.~Townsend,
  ``A Chern-Simons action for three-dimensional anti-de Sitter supergravity
 theories,''
  Phys.\ Lett.\  B {\bf 180}, 89 (1986).

\bibitem{BILS}
I.~A.~Bandos, E.~Ivanov, J.~Lukierski and D.~Sorokin,
``On the superconformal flatness of AdS superspaces,''
JHEP \textbf{06}, 040 (2002) 
[arXiv:hep-th/0205104 [hep-th]].

\bibitem{GGRS}
 S.~J.~Gates, Jr., M.~T.~Grisaru, M.~Ro\v{c}ek and W.~Siegel,
{\it Superspace, or One Thousand and One Lessons in Supersymmetry},
Front.\ Phys.\  {\bf 58}, 1 (1983) [arXiv:hep-th/0108200].



\bibitem{KT-M11} 
  S.~M.~Kuzenko and G.~Tartaglino-Mazzucchelli,
  ``Three-dimensional N=2 (AdS) supergravity and associated supercurrents,''
  JHEP {\bf 1112}, 052 (2011)
  [arXiv:1109.0496 [hep-th]].



\bibitem{Ideas} I.~L.~Buchbinder and S.~M.~Kuzenko,
{\it Ideas and Methods of Supersymmetry and
Supergravity or a Walk Through Superspace}, IOP, Bristol, 1995
(Revised Edition: 1998).



\bibitem{Keck}
  B.~W.~Keck,
 ``An alternative class of supersymmetries,''
J.\ Phys.\ A  {\bf 8}, 1819 (1975).

\bibitem{Zumino77}
B.~Zumino, ``Nonlinear realization of supersymmetry in de Sitter space,''
Nucl.\ Phys.\  B {\bf 127}, 189 (1977).

\bibitem{IS}
E.~A.~Ivanov and A.~S.~Sorin,
``Superfield formulation of OSp(1,4) supersymmetry,''
J.\ Phys.\ A  {\bf 13} (1980) 1159.


\bibitem{KLRT-M1}
S.~M.~Kuzenko, U.~Lindstr\"om, M.~Ro\v cek and G.~Tartaglino-Mazzucchelli,
``4D N=2 supergravity and projective superspace,'' 
JHEP {\bf 0809}, 051 (2008) [arXiv:0805.4683].

\bibitem{KT-M08}
  S.~M.~Kuzenko and G.~Tartaglino-Mazzucchelli,
  ``Field theory in 4D N=2 conformally flat superspace,''
  JHEP {\bf 0810}, 001 (2008)
  [arXiv:0807.3368 [hep-th]].

\bibitem{Butter:2011ym}
D.~Butter and S.~M.~Kuzenko,
``N=2 AdS supergravity and supercurrents,''
JHEP \textbf{07}, 081 (2011)
[arXiv:1104.2153 [hep-th]].

\bibitem{BKLT-M}
D.~Butter, S.~M.~Kuzenko, U.~Lindstr\"om and G.~Tartaglino-Mazzucchelli,
``Extended supersymmetric sigma models in AdS$_4$ from projective superspace,''
JHEP \textbf{05}, 138 (2012)
[arXiv:1203.5001 [hep-th]]

\bibitem{BFP}
D.~J.~Binder, D.~Z.~Freedman and S.~S.~Pufu,
``A bispinor formalism for spinning Witten diagrams,''
[arXiv:2003.07448 [hep-th]].


\bibitem{Siegel93}
  W.~Siegel,
  ``Green-Schwarz formulation of self-dual superstring,''
  Phys.\ Rev.\ D {\bf 47}, 2512 (1993)
  [hep-th/9210008].

\bibitem{Siegel95}
  W.~Siegel,
  ``Super multi-instantons in conformal chiral superspace,''
  Phys.\ Rev.\ D {\bf 52}, 1042 (1995)
  [hep-th/9412011].




\bibitem{Goldberger:2011yp}
  W.~D.~Goldberger, W.~Skiba and M.~Son,
  ``Superembedding methods for 4d N=1 SCFTs,''
  Phys.\ Rev.\ D {\bf 86}, 025019 (2012)
  [arXiv:1112.0325 [hep-th]].


\bibitem{Maio}
  M.~Maio,
  ``Superembedding methods for 4d N-extended SCFTs,''
  Nucl.\ Phys.\ B {\bf 864}, 141 (2012)
  [arXiv:1205.0389 [hep-th]].

\bibitem{Goldberger:2012xb}
  W.~D.~Goldberger, Z.~U.~Khandker, D.~Li and W.~Skiba,
  ``Superembedding methods for current superfields,''
  Phys.\ Rev.\ D {\bf 88}, 125010 (2013)
  [arXiv:1211.3713 [hep-th]].


\bibitem{Sorokin}
  D.~P.~Sorokin,
  ``Superbranes and superembeddings,''
  Phys.\ Rept.\  {\bf 329}, 1 (2000)
  [hep-th/9906142].

\bibitem{Fitzpatrick:2014oza}
  A.~L.~Fitzpatrick, J.~Kaplan, Z.~U.~Khandker, D.~Li, D.~Poland and D.~Simmons-Duffin,
  ``Covariant approaches to superconformal blocks,''
JHEP \textbf{08}, 129 (2014)
[arXiv:1402.1167 [hep-th]].
06142;

\bibitem{Khandker:2014mpa}
  Z.~U.~Khandker, D.~Li, D.~Poland and D.~Simmons-Duffin,
  ``$\mathcal{N}=1$ superconformal blocks for general scalar operators,''
 JHEP \textbf{08}, 049 (2014)
[arXiv:1404.5300 [hep-th]]. 
  

\bibitem{Osborn}
  H.~Osborn,   ``N = 1 superconformal symmetry in four-dimensional
quantum field theory,''
Annals Phys.\  {\bf 272}, 243 (1999) [hep-th/9808041].

\bibitem{Park4}
 J.-H.~Park,  ``Superconformal symmetry and correlation functions,''
 Nucl.\ Phys.\ B {\bf 559}, 455 (1999)   [hep-th/9903230].

\bibitem{KT}
S.~M.~Kuzenko and S.~Theisen,
 ``Correlation functions of conserved currents in N = 2 superconformal
theory,''  Class.\ Quant.\ Grav.\  {\bf 17}, 665 (2000)  [hep-th/9907107].

\bibitem{Park3}
  J.-H.~Park,
  ``Superconformal symmetry in three-dimensions,''
  J.\ Math.\ Phys.\  {\bf 41}, 7129 (2000)
  [arXiv:hep-th/9910199].
  
\bibitem{GIKOS}
  A.~Galperin, E.~Ivanov, S.~Kalitsyn, V.~Ogievetsky and E.~Sokatchev,
  ``Unconstrained N = 2 matter, Yang-Mills and supergravity theories in harmonic
  superspace,''
  Class.\ Quant.\ Grav.\  {\bf 1}, 469 (1984).

\bibitem{GIOS}
A.~S.~Galperin, E.~A.~Ivanov, V.~I.~Ogievetsky and E.~S.~Sokatchev,
{\it Harmonic Superspace}, Cambridge University Press,  2001.

\bibitem{KLR}
A. Karlhede, U. Lindstr\"om and M. Ro\v cek,
``Self-interacting tensor multiplets in N = 2 superspace,''
Phys.\ Lett.\ B {\bf 147}, 297 (1984). 

\bibitem{LR-projective}
U.~Lindstr\"om and M.~Ro\v{c}ek,
``New hyperk\"ahler  metrics  and new supermultiplets,''
  Commun.\ Math.\ Phys.\  {\bf 115}, 21 (1988).
  
\bibitem{LR-projective2}  
U.~Lindstr\"om and M.~Ro\v{c}ek,
 ``N = 2 super Yang-Mills theory in projective superspace,''
Commun.\ Math.\ Phys.\  {\bf 128}, 191 (1990).
  
\bibitem{Costa:2011mg}
M.~S.~Costa, J.~Penedones, D.~Poland and S.~Rychkov,
``Spinning conformal correlators,''
JHEP \textbf{11}, 071 (2011)
[arXiv:1107.3554 [hep-th]].

\end{thebibliography}
\end{document}